\documentclass[11pt,a4paper]{emulateapj}
\usepackage{epsfig}
\usepackage{amsmath}
\usepackage{natbib}

\begin{document}

\title{Correlation between the isotropic energy and the peak energy at  zero fluence for the
 individual pulses of GRBs: towards 
an universal physical correlation for the prompt emission}

\author{Rupal Basak$^{1}$ and A.R. Rao$^{2}$}

\affil{$^{1,2}$Tata Institute of Fundamental Research, Mumbai - 400005, India. 
$^{1}rupalb@tifr.res.in$,
$^{2}arrao@tifr.res.in$}

\begin{abstract}
We find a strong correlation between the peak energy at zero fluence ($\rm E_{peak,0}$) 
and the isotropic energy ($\rm E_{\gamma,iso}$) of the 22 pulses of 9 Gamma Ray 
Bursts (GRB)  detected by the Fermi satellite. The correlation holds for the individual 
pulses of each GRB, which shows the reality of the correlation. The derived correlation 
(Spearman correlation coefficient, $r$, is  0.96)  is much stronger compared to the correlations using $\rm E_{peak}$
(in place of $\rm E_{peak,0}$) determined
from the  time-integrated spectrum ($r$ = 0.8), or the time-resolved spectrum not accounting for
broad pulse structures ($r$ = 0.37), or the pulse-wise spectrum ($r$ = 0.89).  Though the 
improvement in the $\rm E_{peak}$ - $\rm E_{\gamma,iso}$ relation
(the Amati relation) for a pulse-wise analysis is known earlier, this is the first time a parameter derived
from a joint spectral and timing fit to the data is shown to improve the correlation.
We suggest that $\rm E_{peak,0}$, rather than $\rm E_{peak}$, is intrinsic to a GRB pulse and a
natural choice as the parameter in the pulse-wise correlation studies.  
\end{abstract}


\section{INTRODUCTION}
Gamma Ray Bursts (GRBs) are the most luminous events in the universe, predominantly observed in the hard X-ray
to gamma ray energies. An individual GRB appears as a flash of gamma ray event lasting for seconds and then continuously
shifts towards the lower energy bands -- all the way to radio wavelength (see for example, M{\'e}sz{\'a}ros 2006). 
In the initial phase of high
energy emission, known as the prompt emission, it undergoes significant spectral evolution. Over the 
years, researchers have developed empirical models to describe the time-integrated spectrum, the light curve and
the spectral evolution. It has been shown that various model parameters correlate with the energy related 
physical parameters of the full GRB. For example, the parameter peak energy of the $\rm\nu F_{\nu}$ 
spectra ($\rm E_{peak}$) correlates with the isotropic energy ($\rm E_{\gamma,iso}$) (Amati et al. 2002), 
isotropic peak luminosity ($\rm L_{iso}$) (Schaefer 2003; Yonetoku et al. 2004), and collimation-corrected 
energy ($\rm E_{\gamma}$) (Ghirlanda et al. 2004). Similar correlations hold in the time domain e.g., spectral
delay ($\tau_{lag}$) - $\rm L_{iso}$ (Norris, Marani \& Bonnell 2000), variability (V) - $\rm L_{iso}$
(Fenimore \& Ramirez-Ruiz 2000) and rise time ($\rm\tau_{rise}$) - $\rm L_{iso}$ (Schaefer 2007).
The objective of such correlation studies is two-fold. Firstly, to use GRB as a standard candle
of cosmology, alongside the commonly used standard candle -- supernova. Secondly, to understand 
the GRB physics itself. GRB can be observed at a much higher redshift (as high as z $>$ 6) than supernova (z $\le$ 1.7).
With the advent of the Fermi satellite we now have the record holder -- GRB 090423 (z = 8.2; see Ghirlanda et al. 2010).
In spite of this tremendous advantage over the supernova, GRB has one serious drawback.
Unlike the supernova, whose mechanism is
very well understood and supported by numerous observations, the physics of GRB is not understood
properly. Hence, it is interesting to investigate empirical models which can predict the
correct energetics of GRB. The predictions of theoretical model should conform to that of a data driven 
empirical model and this whole process gives a strong constraint on the possible physics of GRB. 

Pulse analysis of GRB prompt emission has received considerable attention in recent
times. It has the potential to unravel physical processes responsible for the
observed  correlations and help in standardizing the energy budget so that GRBs can be used as 
precise cosmological distance indicators (Hakkila 2009). In the prompt emission phase, a GRB 
varies vastly in time and energy domains, so much so that no two bursts have the same temporal and spectral 
characteristics (Norris et al. 1996). This situation simplifies if the whole GRB event is considered 
as an ensemble of temporally separated pulses of self similar shapes (Nemiroff 2000), generated nearly 
simultaneously in a wide range of energy bands (Norris et al. 2005). Hence, each of the pulses can be modelled 
separately and then the whole GRB event can be reproduced by shifting and adding these pulses. 
One of the strongest constraint the pulse analysis gives is that the pulses of a GRB, despite having separate set 
of parameters and different energetics, have the same redshift. Hence, each pulse can be used as
a distance indicator and must conform with each other. It has been shown that all
the correlations studied for the full GRB event hold good, sometimes even better, if the inherent 
pulse property of GRB is taken into account. Krimm et al. (2009), for example, showed that the 
$\rm E_{peak}$ - $\rm E_{\gamma,iso}$ correlation derived from pulse-wise analysis is consistent 
with the time-integrated analysis. Hakkila et al. (2008) studied $\tau_{lag}$ - $\rm L_{iso}$ correlation 
and conclusively showed that the spectral lags are pulse properties rather than burst properties. 

These kind of correlations are, however, empirical in nature, and may not have any physical significance
(Band \& Preece 2005; Butler et al. 2007; 2009). Moreover, they do not use the full information  available 
in the sense that the spectral and timing parameters are considered 
independent of each other. In such correlation studies, the information of time evolution of the spectral parameters
and the energy dependence of the timing parameters are lost. Hence, the derived spectral and timing
parameters are average quantities. Liang \& Kargatis (1996) 
showed that one of the spectral parameters, namely $\rm E_{peak}$, follows a time evolution law.
Kocevski \& Liang (2003) used this interdependence to study the evolution of $\rm E_{peak}$,
and found that $\rm E_{peak}$ exponentially decreases with the fluence of an individual pulse
in a GRB. Recently, Basak \& Rao (2012) have developed a new method for the complete empirical description of the individual 
pulses of GRB prompt emission, simultaneously in the time and energy domains, based on well established empirical 
formulae. The method conclusively shows that the two new parameters of the Liang \& Kargatis (1996) 
model, namely, the peak energy at zero fluence ($\rm E_{peak,0}$) and the characteristic fluence 
($\phi_{0}$), can be used to predict the timing parameters, namely, spectral delay and width. In other words, these two
parameters are more intrinsic to the GRB pulses than the average spectral and timing parameters,
and the other pulse characteristics can be derived from them. In this paper we demonstrate 
that the pulse-wise analysis gives a better correlation than both the time-integrated analysis and 
time-resolved analysis, which do not take broad pulses into account. We also show that $\rm E_{peak,0}$, 
instead of $\rm E_{peak}$, gives even better correlation, and it is the correct choice of parameter in correlation studies. 
The structure of the paper is as follows. In \S 2 we briefly describe the method and in \S 3 we give 
the data analysis and results. Major conclusions are discussed in \S 4.

\section{Simultaneous timing and spectral description of a GRB pulse}
Pulse analysis of GRB essentially involves extracting the constituent substructures based on
various empirical time description of pulses (Norris et al. 1996; Nemiroff et al. 2000; Hakkila 
et al. 2009; 2011), e.g., the fast rise exponential decay model (FRED; Kocevski 2003), the Norris 
model (Norris et al. 2005). Though this method is useful in extracting isolated and
slightly overlapping pulses, it cannot be used to extract ``heavily-overlapping'' and
low signal-to-noise pulses at present (Hakkila et al. 2011). The energy spectrum of a GRB is popularly 
described by an empirical model given by Band et al. (1993). The same model can also be used for the 
individual pulses. Choosing Norris model for the time domain and the Band model for the
energy domain, and employing the time evolution of the peak energy ($\rm E_{peak}$)
of Band model, as proposed by Liang \& Kargatis (1996), Basak \& Rao (2012) developed
a technique to determine the model parameters ($\rm E_{peak,0}$ and $\phi_{0}$). 

Basak \& Rao (2012) analyzed the Swift/BAT and Fermi/GBM data of the brightest GRB in the 
Fermi era -- GRB 090618. They measured the global parameters of the individual pulses of this GRB, 
namely, the low energy photon index ($\alpha$), high energy photon index ($\beta$) using the Band 
model (Band et al. 1993) for the time-integrated spectral data, and the characteristic time scales 
($\tau_1$ and $\tau_2$) by fitting the energy-integrated light curve with the Norris 
model (Norris et al. 2005). These global parameters were then used to generate the XSPEC table model 
with the parameters $\rm E_{peak,0}$ and $\phi_{0}$ as variables. From spectral fitting in XSPEC, $\rm E_{peak,0}$ 
and $\phi_{0}$ were determined.

The data analysis for the global parameters (see Basak \& Rao 2012) showed improvement due to the inclusion of
the Swift/BAT along with the Fermi/GBM. The simultaneous spectral fitting, however,
showed some systematic errors in the overlapping energy regions. Hence, 
they used only Swift/BAT data for the simultaneous timing and spectral description of the GRB pulses. 
The typical value of the peak energy ($\rm E_{peak}$) of GRBs (and the individual pulses) is 
$\sim$ 300 keV, and we expect still higher values of the parameter, $\rm E_{peak,0}$. 
The Swift/BAT energy range (15 - 150 keV; Barthelmy et al. 2005)
is inadequate for accurate measurement of these parameters in many occasions. Hence, the
derived values by Basak and Rao (2012) showed large uncertainties (typically 50 -- 100 keV). 
GBM onboard the
Fermi satellite provides an unprecedented energy range ($\sim$8 keV to 
$\sim$40 MeV) with adequate sensitivity (Meegan 2009; Meegan et al. 2009) for
constraining the values of $\rm E_{peak}$ and $\rm E_{peak,0}$. 
Hence, in the present analysis, we have used Fermi/GBM data. Also, we have carefully chosen the time
interval of each pulse avoiding contamination from the other pulses. For example,
in GRB 090618 (see Table 1), we have chosen time intervals 61-76, 76-95 and 106-126 seconds
for the pulses 2, 3 and 4 respectively (compared with the time divisions of Basak \& Rao (2012): 
50-77, 77-100 and 100-130 seconds, respectively). We have included the precursor burst of this GRB (pulse 1) in the
present analysis. 
\begin{table*}[ht]\centering
\caption{The observer frame peak energy ($E_{peak}$), zero fluence peak
energy ($E_{peak,0}$) and the isotropic energy ($E_{\gamma,iso}$)
for the individual pulses of GRBs (GRB data set taken from Ghirlanda
et al. 2010) }

\begin{tabular}{cccccccccc}
\hline 
GRB & z & Pulse & $t_{1}$(s) & $t_{2}$(s) & $E_{peak}$(keV) & $\chi_{red}^{2}$ & $E_{peak,0}$(keV) & $\chi_{red}^{2}$ & $E_{\gamma,iso}$($10^{52}$erg)\tabularnewline
\hline
\hline 
080810 & 3.35 & 1 & 20.0 & 28.0 & $354_{-61}^{+188}$ & 0.95 & $875_{-180}^{+155}$ & 0.99 & 7.7\tabularnewline
\hline 
080916C & 4.35 & 1 & 0.0 & 13.0 & $430_{-67}^{+87}$ & 1.17 & $2420_{-397}^{+523}$ & 1.21 & 158.9\tabularnewline
 &  & 2 & 16.0 & 43.0 & $477_{-82}^{+108}$ & 1.11 & $1575_{-150}^{+170}$ & 1.38 & 130.1\tabularnewline
\hline 
080916 & 0.689 & 1 & -1.0 & 10.0 & $155_{-19}^{+23}$ & 1.12 & $519_{-59}^{+44}$ & 1.36 & 0.78\tabularnewline
 &  & 2 & 13.0 & 25.0 & $70_{-9}^{+13}$ & 1.0 & $226_{-28}^{+285}$ & 1.00 & 0.25\tabularnewline
 &  & 3 & 28.0 & 39.0 & $39.7_{-7}^{+9}$ & 1.09 & $70_{-23}^{+12}$ & 1.10 & 0.05\tabularnewline
\hline 
081222 & 2.77 & 1 & -2.0 & 20.0 & $159_{-17}^{+22}$ & 1.42 & $488_{-156}^{+173}$ & 1.07 & 23.7\tabularnewline
\hline 
090323 & 3.57 & 1 & -2.0 & 30.0 & $697_{-51}^{+51}$ & 1.33 & $2247_{-298}^{+392}$ & 1.46 & 127.9\tabularnewline
 &  & 2 & 59.0 & 74.0 & $476_{-47}^{+57}$ & 1.38 & $1600_{-94}^{+35}$ & 1.95 & 90.3\tabularnewline
 &  & 3 & 137.0 & 150.0 & $117_{-28}^{+31}$ & 1.37 & $211_{-63}^{+54}$ & 1.17 & 20.9\tabularnewline
\hline 
090328 & 0.736 & 1 & 3.0 & 9.0 & $648_{-124}^{170}$ & 0.93 & $1234_{-146}^{+174}$ & 0.92 & 2.8\tabularnewline
 &  & 2 & 9.0 & 20.0 & $659_{-106}^{+115}$ & 1.25 & $1726_{-122}^{+221}$ & 1.41 & 4.4\tabularnewline
 &  & 3 & 55.0 & 68.0 & $89_{-20}^{+41}$ & 1.16 & $180_{-96}^{+267}$ & 1.03 & 0.36\tabularnewline
\hline 
090423 & 8.2 & 1 & -11.0 & 13.0 & $76.9_{-26}^{+56}$ & 1.10 & $131_{-43}^{+99}$ & 1.04 & 20.3\tabularnewline
\hline 
090424 & 0.544 & 1 & -0.5 & 3.0 & $153_{-5}^{+6}$ & 1.73 & $184.5_{-19}^{+38}$ & 1.43 & 2.0\tabularnewline
 &  & 2 & 3.0 & 6.0 & $148_{-7}^{+8}$ & 1.41 & $162_{-9.2}^{+61}$ & 1.32 & 1.4\tabularnewline
 &  & 3 & 6.5 & 13.0 & $39.1_{-8.4}^{+0.2}$ & 1.32 & $104.8_{-16}^{+17}$ & 1.27 & 0.18\tabularnewline
 &  & 4 & 13.5 & 20.0 & $19.6_{-14.8}^{+6.4}$ & 0.98 & $75_{-34}^{+23}$ & 0.95 & 0.10\tabularnewline
\hline 
090618 & 0.54 & 1 & -1.0 & 41.0 & $185_{-25}^{+26}$ & 1.24 & $415_{-28}^{+37}$ & 1.19 & 3.5\tabularnewline
 &  & 2 & 61.0 & 76.0 & $226_{-9}^{+10}$ & 1.25 & $382_{-30}^{+106}$ & 1.33 & 9.8\tabularnewline
 &  & 3 & 76.0 & 95.0 & $128_{-5}^{+6}$ & 1.15 & $218_{-5.8}^{+6.8}$ & 1.09 & 5.4\tabularnewline
 &  & 4 & 106.0 & 126.0 & $57.7_{-3.3}^{+3.5}$ & 1.08 & $205_{-12}^{+14}$ & 1.19 & 1.5\tabularnewline
\hline
\end{tabular}
\end{table*}

\section{Data Analysis and Results}
In our analysis, we have chosen the set of GRBs which were considered for time-resolved spectral analysis
by Ghirlanda et al. (2010). This contains 12 long GRBs with known redshift, detected by Fermi/GBM 
till the end of July 2009. Among these, three GRBs (GRB 080905, GRB 080928, and GRB 081007)
are very weak bursts and could be fit only with a single power law with unconstrained peak energy. 
Hence, we have used the data for the remaining 9 GRBs and their individual pulses (a total of 22 pulses).
Table 1 contains the full list of our sample with the name, measured redshift, and pulses of GRBs in 
the first, second, and the third columns, respectively.

In a given GRB, we select those pulses which have broader width compared to the rapid spikes in its light curve. 
While selecting such pulses we carefully avoid those pulses or portions of the pulses 
which have too much overlap with others. Fishman et al. (1995), for example, 
have discussed various categories of light curves: 1. single pulse, 2. smooth, either single or multiple, 
well-defined peaks, 3. distinct, well-separated episodes of emission, 4. very erratic, chaotic and spiky 
burst. More than one such category of pulses can show up in a single GRB. In such cases, the fourth category gives
rise to many overlapping pulses in some parts, while the other parts of the same GRB might be dominated by the 
second or third category of pulses. The temporal regions populated by the fourth category do not allow a unique 
measurement of model parameters of a pulse, if each pulse has to have an independent set of parameters. 
Hence, we avoid such regions and take only clear portions of a burst. 

The broad pulses selected in our analysis are listed in Table 1 (also see Ghirlanda et al. 2010: Figure 2),
along with the appropriate start time ($t_1$) and stop time ($t_2$) shown in columns 4 and 5. 
Ghirlanda et al. (2010) have done a time-resolved analysis, whereas  we have done a pulse-wise
analysis. In the former case, the GRB light curve is arbitrarily divided into a large
number of bins and the evolution of spectra is examined. Krimm et al. (2009) have shown that
the correlation between the peak energy and the isotropic energy of broad GRB pulses
improves from that of the whole GRBs. In our analysis, we follow the same approach and 
select the time-cut according to the broad pulse structure.
Some portions of the light curves are neglected, because, either these are dominated by the fourth category of pulses or
have low count rates to accurately determine model parameters. GRB 080810 and GRB 080916C have
low count rates after 30 s and 55 s respectively. GRB 090323 contains multiple overlapping spikes between 30 to 59 s and
75 to 135 s region which do not contain any broad pulse structure (also count rate is low in 75-135 s region).
GRB 090328 has two overlapping spikes between 20 to 26 s (taken as 20-24 s and 24-26 s in Ghirlanda et al. 2010)
and hence neglected.


We essentially use the method described in Basak \& Rao (2012). The global parameters 
($\alpha$, $\beta$, $\tau_1$ and $\tau_2$) are determined to generate three dimensional
pulse model for a set of $\rm E_{peak,0}$ and $\phi_{0}$. The time-integrated spectra
for these set of values give a two-parameter XSPEC table model. We perform $\chi^{2}$
minimization of the spectral fit of the data with this model to determine the best fit 
values of the model parameters and their nominal 90\% confidence level errors ($\Delta \chi^2$ = 2.7).
Also, the normalization, which is a free parameter in the model, is determined by XSPEC.

The results of our spectral analysis are shown in Table 1. 
The observer frame peak energy ($\rm E_{peak}$) calculated from the Band model, the zero fluence peak 
energy ($\rm E_{peak,0}$) and the measured nominal 90\% confidence level errors for each are shown 
in columns 6 and 8. In order to compare the improvement in correlation
of these parameters with $\rm E_{\gamma,iso}$ (column 10 of Table 1), we perform a linear fit 
of the form $\rm log~(y)=K+\delta~log~(x)$ using the technique of joint likelihood for the coefficient
K, $\delta$ and the intrinsic scatter ($\sigma_{int}$) (D'Agostini 2005; Wang et al. 2011).
Here, $\rm y=E_{peak}$ (full and pulse-wise study) and $\rm E_{peak,0}$, are  in the units of 100 keV, and
$\rm x=E_{\gamma,iso}$ is in the units of 10$^{52}$ ergs. 
 
\begin{figure}[ht]
\includegraphics[scale=0.35]{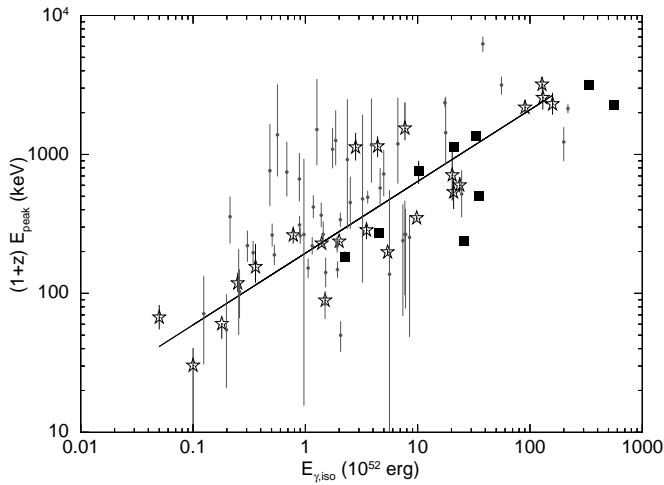}\caption{Time-integrated, time-resolved 
(not accounting for broad pulse structure) and pulse-wise $E_{peak}$ as a function of isotropic
energy ($E_{\gamma,iso}$) for the 9 Fermi-GRBs with measured redshift. The large filled boxes denote the time-integrated points
while small circles denote the time-resolved points (GRBs are chosen from Ghirlanda
et al. 2010). The stars denote the pulse-wise $E_{peak}$ as determined by
the Band model (present work). }

\end{figure}

In Figure 1, we have plotted the 
source peak energy ($\rm E_{peak}$)  as a  function 
of $\rm E_{\gamma,iso}$ for the 22 individual pulses of the 9 GRBs (stars). For comparison, in
the same figure we have plotted the time-integrated values for the full GRB  (filled boxes), 
based on the data given in Ghirlanda et al. (2010). For illustration purposes, a scatter plot
of the values for time-resolved spectral analysis is also given in the figure  (small circles).
Note that L$_{iso}$ is the more appropriate parameter for a time-resolved spectral analysis, but to
compare with our results we have converted the flux given by Ghirlanda et al. (2010) to
  $\rm E_{\gamma,iso}$ using the multiplication factor of the time bin.
The straight line shown in the  figure is the $\rm log~(E_{peak})
=K+\delta~log~(E_{\gamma,iso})$ fit to the pulse-wise data using the joint likelihood method. The time-integrated 
data is also fit by the same technique. The Spearman correlation coefficient (r), probability that the correlation
occurred by chance (P), the parameters for the linear fit and the intrinsic scatter in the data are given in Table 2. 
For comparison, the values reported by Ghirlanda et al. (2010) for the time-integrated values for 10 GRBs 
are also given in the table. A comparison of correlation and P for the time-integrated (0.80 and 
9.60$\times10^{-3}$, respectively) and pulse-wise (0.89 and 2.95$\times10^{-8}$, respectively) analysis 
shows that there is an improvement in the correlation and the reality of the correlation (i.e., lower P) in the latter 
case, favouring pulse-wise analysis. Intrinsic scatter of the data ($\sigma_{int}$) per point is also 
reduced (0.225/9 to 0.244/22). This improvement in the correlation of $\rm E_{peak}$ - $\rm E_{\gamma,iso}$ 
for pulse-wise analysis is known in earlier works. Krimm et al. (2009), for example, used a sample of 
Swift-Suzaku GRBs which gave the Spearman correlation as 0.74 with a chance probability of 7.58$\times 10^{-5}$ 
for a sample of 22 GRBs. This correlation improves, for a set of 59 pulses of these GRBs, to 0.80 with 
a chance probability of 5.32$\times 10^{-14}$. 

The time-resolved study not accounting for broad pulses shows a very poor 
correlation (r=0.37; see Table 2). This is expected because, in such time divisions, broad pulse structure is not 
considered. Even if the spikes are considered as pulses, they are overlapping in any such small time 
division and hence not usable to determine $\rm E_{peak}$, which can be uniquely associated with a pulse. Time divisions taken for 
a broad pulse, on the other hand, has the facility to study the evolution of $\rm E_{peak}$ and uniquely 
determine the $\rm E_{peak,0}$ in that pulse. 
In our analysis, we ignore these rapidly varying spikes and concentrate on the
broad pulse structures.

\begin{figure}
\includegraphics[scale=0.35]{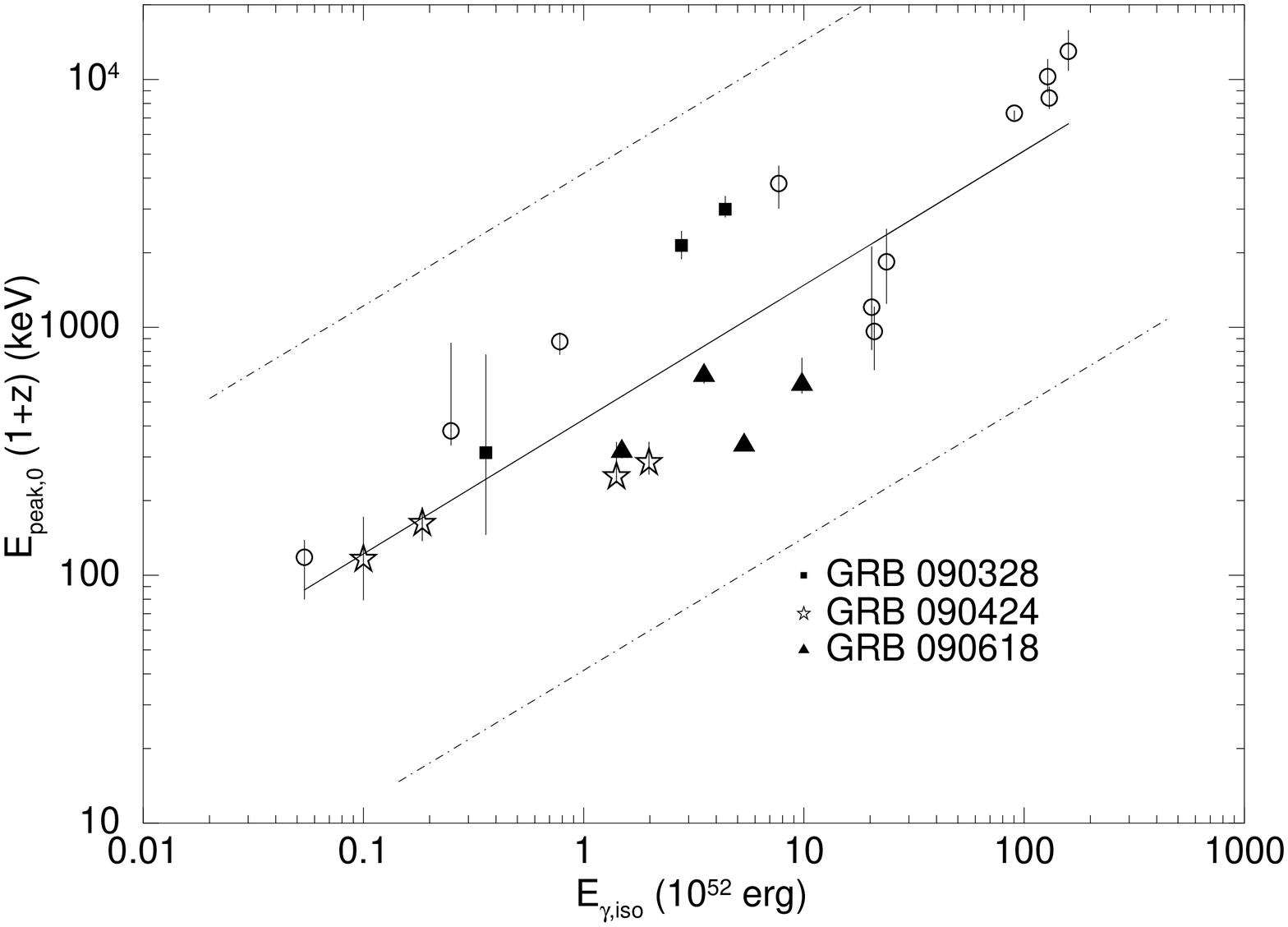}\caption{Pulse-wise $E_{peak,0}$ as a function of isotropic energy ($E_{\gamma,iso}$)
for the 22 pulses of 9 Fermi-GRBs. Data for three individual GRBs (GRB 090328, GRB 090424, and GRB 090618) are
marked with separate symbols. The solid line is the best linear fit, while the dot-dashed lines show the 3$\sigma$
scatter.}

\end{figure}

In Figure 2, we have shown $\rm E_{peak,0}$ as a function of $\rm E_{\gamma,iso}$ along with a
straight line fit. A comparison of Figure 1 with Figure 2 shows immediately that the new parameter
$\rm E_{peak,0}$ has a better correlation in the $\rm E_{peak,0}$ - $\rm E_{\gamma,iso}$ plane compared to
the pulse-wise $\rm E_{peak}$ - $\rm E_{\gamma,iso}$ analysis, which already shows improvement
in terms of correlation, P and $\sigma_{int}$ compared to the time-integrated analysis. All these values 
for the $\rm E_{peak,0}$ - $\rm E_{\gamma,iso}$ fitting are shown in the last row of Table 2. 
The correlation coefficient (r = 0.96) is significantly higher than that of $\rm E_{peak}$ - $\rm E_{\gamma,iso}$,
whether time-integrated (0.80) or  time-resolved (0.37) or pulse-wise (0.89). 
The chance probability also decreases. As a comparison of correlations found in previous works, we quote 
some of the results presented by Krimm et al. (2009): The original Amati (Amati 2006) catalogue of 39 bursts 
has r = 0.87 with chance probability 4.72 $\times 10^{-13}$. A sample of all the long bursts (91) shows
a poor correlation (r = 0.76) with chance probability 4.72 $\times 10^{-18}$. The sample of 59 GRB pulses
of Krimm et al. (2009) has a correlation of 0.80 with a chance probability of 5.32$\times 10^{-14}$. Our analysis shows a
clear improvement in the correlations, whether time-integrated or time-resolved (not accounting for broad
pulse) or pulse-wise analysis. 

\begin{table*}[ht]\centering

\caption{Statistical analysis for the correlations of $E_{peak}$ (I to IV) or $E_{peak,0}$ (V) with $E_{\gamma,iso}$ and parameters 
for the linear fit for the 9 Fermi GRBs and their individual pulses. The
methods are, I: time-integrated study (present work) for 9 GRBs, II: time-integrated study for 10 GRBs
(quoted from Ghirlanda et al. 2010). III: time-resolved study not accounting for broad pulse
structures (calculated from Ghirlanda et al. 2010). IV: pulse-wise study (present work). 
V: Pulse-wise $E_{peak,0}$ - $E_{\gamma,iso}$ correlationThe Spearman correlation coefficient ($r$), 
chance probability (P) and the parameters of linear fit (K, $\delta$ and $\sigma_{int}$) are reported here.}

\begin{tabular}{ccccccc}
\hline 
Method & r & P & K & $\delta$ & $\sigma_{int}$ & $\chi_{red}^{2}$(dof)\tabularnewline
\hline
\hline 
I & 0.80 & 0.0096 & 0.166$\pm$0.080 & 0.473$\pm$0.048 & 0.225$\pm$0.067 & 0.64 (7)\tabularnewline

II & --- & 0.004 & 0.162$\pm$0.085 & 0.476$\pm$0.079 & --- & 0.47 (8)\tabularnewline

III & 0.37 & 0.0095 & --- & --- & --- & ---\tabularnewline

IV & 0.89 & 2.95$\times10^{-8}$ & 0.289$\pm$0.055 & 0.516$\pm$0.049 & 0.244$\pm$0.048 & 0.56 (20)\tabularnewline

V & 0.96 & 1.60$\times10^{-12}$ & 0.640$\pm$0.050 & 0.555$\pm$0.050 & 0.291$\pm$0.039 & 1.04 (20)\tabularnewline
\hline
\end{tabular}
\end{table*}

It is long debated that the correlation such as $\rm E_{peak}$ - $\rm E_{\gamma,iso}$  might be 
a result of observational selection effects. In particular, the $\rm E_{peak}$ values are stretched
due to the multiplication factor (1+z) making the correlation look better (Nakar and Piran 2005;
Band and Preece 2005; Butler et al. 2007, 2009; Shahmoradi and Nemiroff 2009). But, there are
other researchers who argue in favour of the reality of these correlations (Ghirlanda et al. 2008;
Nava et al. 2008; Krimm et al. 2009; Amati et al. 2009; Ghirlanda et al. 2010). If these 
correlations hold within a GRB as for a sample of many GRBs, then we can conclude that the
correlations are indeed physical. In Figure 2, we have marked the data points for GRB 090424 (stars) and GRB 090618 (filled
triangle) for which we analyze the highest number (four) of pulses. The data shows that the 
pulses of the same GRB follow the same $\rm E_{peak,0}$ - $\rm E_{\gamma,iso}$ correlation.
Also, the fact that this correlation is tighter than $\rm E_{peak}$ - $\rm E_{\gamma,iso}$  correlation
signifies that $\rm E_{peak,0}$ is more intrinsic of a GRB pulse than $\rm E_{peak}$ and
hence this parameter should be used for such correlation studies.
It is interesting to note that the pre-cursor of the GRB 0901618 also follows the same correlation. 
The GRB 090328 is also marked (filled squares). First two pulses of this GRB (with higher $\rm E_{peak,0}$ values) 
show particularly high deviation. An examination of the $\rm E_{peak}$ - $\rm E_{\gamma,iso}$  values 
of these two pulses (Figure 1) also shows a much higher deviation. Interestingly, the third pulse of this 
GRB is consistent with the trend. It is known earlier that the first pulses of GRBs tend
to be harder and thus deviate from the correlation (Krimm et al. 2009, Ghirlanda et al. 2010). But,
this deviation is much less compared to the pulse-wise $\rm E_{peak}$ in the $\rm E_{peak}$ - $\rm E_{\gamma,iso}$ 
correlation. 

\section{Discussion and Conclusions}

The strong correlation between the luminosity and the parameter obtained from a joint spectral
and timing fit for the individual pulses of GRBs clearly indicates that the basic radiation/ emission
process is similar  in diverse GRBs and the dispersion in the other parameters like Lorentz factor,
beaming angle etc. are quite minimal. An examination of the data in Table 1 and Figure 2 shows that
the dispersion in the correlation for different pulses of a GRB is of the same magnitude as the
dispersion between GRBs. Hence it is worthwhile to investigate the 
radiation/ emission process in operation by making a direct model fit to the data. 
This would help us finding the fundamental parameter/ variable responsible
for the correlation. If such an exercise brings down the dispersion in the relation, 
it may be possible 
to use GRBs as a 
distance indicator for cosmological purposes. It is also interesting to note that the parameters
derived for the ``pre-cursor'' in GRB~090618, is consistent with the  global correlation.

Asano \& Meszaros (2011) have done simulations of the spectral  and temporal 
evolution of gamma-ray bursts using internal dissipation models and have concluded that
the models reproduce the Band spectra and also the generic time evolution. 
Dado et al. (2007) have used  the master formula 
based on the ``cannonball'' model and have explained the various correlations
observed for the prompt emission of the GRBs. 
The present work
demonstrates that it is possible to fit the data with a comprehensive set of 
formulae describing the temporal and spectral evolution of the 
bursts. Hence, it should be possible to directly fit the data with the model
predictions  and derive the fundamental quantities responsible
for the universal correlation. 
Any model contains basic physical
assumptions along with other details.
A direct fit to the data taking individual pulses should segregate the basics from
the details. For example, in the   ``cannonball'' model (Dado et al. 2007) the Lorentz
factor and the viewing angle of a GRB determine most of the properties of 
a GRB pulse. Since the viewing angle would be the same for the different pulses
of a given GRB, a direct fit to the spectral and temporal profile can have the
additional constraint of the constancy of this parameter.

 Further improvements for the method described here (also see Basak \& Rao (2012)) could
be made by iteratively including color corrections to the light curve and
also  re-confirming
the $\rm E_{peak}$ evolution formula. It would also be interesting to  repeat this exercise for
short GRBs and X-ray flares, which will give further clues to the emission mechanisms 
responsible for the correlation presented here.

The choice of isotropic energy ($\rm E_{\gamma,iso}$) over the other two physical quantities, namely, 
the isotropic peak luminosity ($\rm L_{iso}$) and collimation-corrected energy ($\rm E_{\gamma}$)
can be justified as follows. Peak luminosity is measured based on the assumption that
the spectral shape at the peak is the same as the average shape, which is not
very physical (Wang et al. 2011). On the other hand, $\rm E_{\gamma}$ could have been
a better choice, as collimation effect is corrected. Though, in practice the beaming angle
($\theta_j$) is very ill determined (e.g., see the weak constraints in the measured $\theta_j$
by Goldstein et al. 2011). It is evident from Table 2 that the intrinsic 
scatter does not improve in the new analysis. This might happen due to some
inherent assumptions of our model, crucially the hard-to-soft evolution and/or
the uncertainty in the measured redshift. A close inspection of Figure 2 reveals that
the pulses of the same GRB are scattered on the same side of the correlation line. This might be 
due to the fact that the correct energy budget is the collimation corrected energy and
not the isotropic energy. Analysis of 14 pulses of 6 GRBs (for which $\theta_j$
values could be collected) shows a correlation of 0.91. We believe that this correlation
might improve with increasing accuracy of the measured $\theta_j$.

To summarize, we have used the method of joint timing and spectral description of
GRB pulses and found that $\rm E_{peak,0}$ is a fundamental parameter in the pulse
description.

\section*{Acknowledgements} This research has made use of data obtained through the
HEASARC Online Service, provided by the NASA/GSFC, in support of NASA High Energy
Astrophysics Programs. We are greatly thankful to the referee for the valuable
suggestions in many places which improved the readability of the paper.


\begin{thebibliography}{}

\bibitem[Amati(2006)]{2006MNRAS.372..233A} Amati, L.\ 2006, \mnras, 372, 
233 


\bibitem[Amati et 
al.(2002)]{2002A&A...390...81A} Amati, L., Frontera, F., Tavani, M., et al.\ 2002, \aap, 390, 81 

\bibitem[Amati et 
al.(2009)]{2009A&A...508..173A} Amati, L., Frontera, F., \& Guidorzi, C.\ 2009, \aap, 508, 173 



\bibitem[Asano (2011)]{2011} Asano, K. \& M{\'e}sz{\'a}ros, P.\ 
2011, arXiv:1107.4825v2 

\bibitem[Band et al.(1993)]{1993ApJ...413..281B} Band, D., Matteson, J., 
Ford, L., et al.\ 1993, \apj, 413, 281 


\bibitem[Band 
\& Preece(2005)]{2005ApJ...627..319B} Band, D.~L., \& Preece, R.~D.\ 2005, \apj, 627, 319 

\bibitem[Barthelmy et al.(2005)]{2005SSRv..120..143B} Barthelmy, S.~D., 
Barbier, L.~M., Cummings, J.~R., et al.\ 2005, \ssr, 120, 143 

\bibitem[Basak 
\& Rao(2012)]{2012ApJ...745...76B} Basak, R., \& Rao, A.~R.\ 2012, \apj, 745, 76 



\bibitem[Butler et al.(2009)]{2009ApJ...694...76B} Butler, N.~R., Kocevski, 
D., \& Bloom, J.~S.\ 2009, \apj, 694, 76 


\bibitem[Butler et al.(2007)]{2007ApJ...671..656B} Butler, N.~R., Kocevski, 
D., Bloom, J.~S., \& Curtis, J.~L.\ 2007, \apj, 671, 656 

\bibitem[Dado et al.(2007)]{2007ApJ...663..400D} Dado, S., Dar, A., 
\& De R{\'u}jula, A.\ 2007, \apj, 663, 400 


\bibitem[D'Agostini(2005)]{2005foap.conf...79D} D'Agostini, G.\ 2005, 
Frontier Objects in Astrophysics and Particle Physics, Vulcano Workshop 
2004, held 24-29 May 2004 in Vulcano, Italy.~Edited by F.~Giovannelli and 
G.~Mannocchi, 2005, p.79, 79 

\bibitem[Fenimore 
\& Ramirez-Ruiz(2000)]{2000astro.ph..4176F} Fenimore, E.~E., \& Ramirez-Ruiz, E.\ 2000, arXiv:astro-ph/0004176 

\bibitem[Fishman 
\& Meegan(1995)]{1995ARA&A..33..415F} Fishman, G.~J., \& Meegan, C.~A.\ 1995, \araa, 33, 415 














\bibitem[Ghirlanda et 
al.(2010)]{2010A&A...511A..43G} Ghirlanda, G., Nava, L., \& Ghisellini, G.\ 2010, \aap, 511, A43 

\bibitem[Ghirlanda et al.(2008)]{2008MNRAS.387..319G} Ghirlanda, G., Nava, 
L., Ghisellini, G., Firmani, C., \& Cabrera, J.~I.\ 2008, \mnras, 387, 319 

\bibitem[Ghirlanda et al.(2004)]{2004ApJ...616..331G} Ghirlanda, G., 
Ghisellini, G., \& Lazzati, D.\ 2004, \apj, 616, 331 

\bibitem[Goldstein et al.(2011)]{2011arXiv:1101.2458v1}Goldsein, A. et al., 2011, arXiv:1101.2458v1

\bibitem[Hakkila 
\& Preece(2011)]{2011arXiv1103.5434H} Hakkila, J., \& Preece, R.~D.\ 2011, arXiv:1103.5434 


\bibitem[Hakkila et al.(2009)]{2009AIPC.1133..479H} Hakkila, J., Fragile, 
P.~C., 
\& Giblin, T.~W.\ 2009, American Institute of Physics Conference Series, 1133, 479 


\bibitem[Hakkila et al.(2008)]{2008ApJ...677L..81H} Hakkila, J., Giblin, 
T.~W., Norris, J.~P., Fragile, P.~C., 
\& Bonnell, J.~T.\ 2008, \apjl, 677, L81 

\bibitem[Kocevski et al.(2003)]{2003ApJ...596..389K} Kocevski, D., Ryde, 
F., \& Liang, E.\ 2003, \apj, 596, 389 


\bibitem[Kocevski 
\& Liang(2003)]{2003ApJ...594..385K} Kocevski, D., \& Liang, E.\ 2003, \apj, 594, 385 

\bibitem[Krimm et al.(2009)]{2009ApJ...704.1405K} Krimm, H.~A., Yamaoka, 
K., Sugita, S., et al.\ 2009, \apj, 704, 1405 



\bibitem[Liang 
\& Kargatis(1996)]{1996Natur.381...49L} Liang, E., \& Kargatis, V.\ 1996, \nat, 381, 49 






\bibitem[M{\'e}sz{\'a}ros(2006)]{2006RPPh...69.2259M} M{\'e}sz{\'a}ros, P.\ 
2006, Reports on Progress in Physics, 69, 2259 






\bibitem[Meegan(2009)]{2009APS..APR.T4001M} Meegan, C.\ 2009, APS April 
Meeting Abstracts, 4001 


\bibitem[Meegan et al.(2009)]{2009ApJ...702..791M} Meegan, C., Lichti, G., 
Bhat, P.~N., et al.\ 2009, \apj, 702, 791 

\bibitem[Nakar 
\& Piran(2005)]{2005MNRAS.360L..73N} Nakar, E., \& Piran, T.\ 2005, \mnras, 360, L73 





\bibitem[Nava et al.(2008)]{2008MNRAS.391..639N} Nava, L., Ghirlanda, G., 
Ghisellini, G., \& Firmani, C.\ 2008, \mnras, 391, 639 


\bibitem[Nemiroff(2000)]{2000ApJ...544..805N} Nemiroff, R.~J.\ 2000, \apj, 
544, 805 

\bibitem[Norris et al.(2005)]{2005ApJ...627..324N} Norris, J.~P., Bonnell, 
J.~T., Kazanas, D., et al.\ 2005, \apj, 627, 324 


\bibitem[Norris et al.(2000)]{2000ApJ...534..248N} Norris, J.~P., Marani, 
G.~F., \& Bonnell, J.~T.\ 2000, \apj, 534, 248 


\bibitem[Norris et al.(1996)]{1996ApJ...459..393N} Norris, J.~P., Nemiroff, 
R.~J., Bonnell, J.~T., et al.\ 1996, \apj, 459, 393 









\bibitem[Schaefer(2007)]{2007ApJ...660...16S} Schaefer, B.~E.\ 2007, \apj, 
660, 16 


\bibitem[Schaefer(2003)]{2003ApJ...583L..67S} Schaefer, B.~E.\ 2003, \apjl, 
583, L67 

\bibitem[Shahmoradi 
\& Nemiroff(2009)]{2009AIPC.1133..425S} Shahmoradi, A., \& Nemiroff, R.\ 2009, American Institute of Physics Conference Series, 1133, 425 








\bibitem[Wang et al.(2011)]{2011MNRAS.415.3423W} Wang, F.-Y., Qi, S., 
\& Dai, Z.-G.\ 2011, \mnras, 415, 3423 






\bibitem[Yonetoku et al.(2004)]{2004ApJ...609..935Y} Yonetoku, D., 
Murakami, T., Nakamura, T., et al.\ 2004, \apj, 609, 935 






\end{thebibliography}
\end{document}